\documentclass[12pt]{article}   
\usepackage{latexsym}
\usepackage{amsmath}      
\usepackage{amsfonts}
\usepackage{amssymb}      
\usepackage{mathrsfs}       
\usepackage{bm}              
\usepackage{graphicx,epic,eepic}      
\usepackage{graphics}      
\usepackage{epsfig}
\usepackage[dvips]{color}
\usepackage{cite}             
\usepackage{hyperref}
\makeindex
\begin{document}
\title{%
{\Huge \bf The Kerr spacetime:}\\[10pt]
{\Huge \bf  A brief introduction}\\[15pt]
}
\author{%
{\bf Matt Visser}\\
School of Mathematics, Statistics, and Computer Science\\
Victoria University of Wellington\\
PO Box 600\\
Wellington\\
New Zealand
}
\date{5 June 2007;  revised 30 June 2007; revised 15 January 2008; \LaTeX-ed \today}
\maketitle

Comment: This is a draft of an introductory chapter on the Kerr spacetime that is intended for use in the book {\bf ``The Kerr spacetime''}, currently being edited by Susan Scott, Matt Visser, and David Wiltshire. This chapter is intended as a teaser and brief introduction to the mathematics and physics of the Kerr spacetime --- it is not, nor is it intended to be, a complete and exhaustive survey of everything in the field.
Comments and community feedback, especially regarding clarity and pedagogy,  are welcome.

\vskip 5pt 


\vskip 5pt 

arXiv:0706.0622 [gr-qc]

\vskip 10pt



\begin{abstract}
This chapter provides a brief introduction to the mathematics and physics of the Kerr spacetime and rotating black holes, touching on the most common coordinate representations of the spacetime metric and the key features of the geometry --- the presence of horizons and ergospheres. The coverage is by no means complete, and serves chiefly to orient oneself when reading subsequent chapters.
\end{abstract}

\def\Box{\diamondsuit} 

\def\eof{\Box}

\def\d{{\mathrm{d}}}
\def\implies{\Rightarrow}
\def\ie{{\emph{i.e.}}}
\def\eg{{\emph{e.g.}}}
\def\etc{{\emph{etc.}}}

\def\real{I\!\!R}
\def\R{I\!\!R}
\def\nat{I\!\!N}

\def\lsim{\mathrel{\lower2.5pt\vbox{\lineskip=0pt\baselineskip=0pt
          \hbox{$<$}\hbox{$\sim$}}}}
\def\gsim{\mathrel{\lower2.5pt\vbox{\lineskip=0pt\baselineskip=0pt
          \hbox{$>$}\hbox{$\sim$}}}}
\def\real{\mathrel{\lower.0pt \hbox{$I\!\!R$}}}
\def\lint{\hbox{\Large $\displaystyle\int$}} 
\def\hint{\hbox{\Huge $\displaystyle\int$}}  

\label{C:Introduction}

\clearpage
\section{Background}

The Kerr spacetime has now been with us for some 45 years\cite{Kerr,Kerr-Texas}. It was discovered in 1963 through an intellectual \emph{tour de force}, and continues to provide highly nontrivial and challenging mathematical and physical problems to this day. 

The final form of Albert Einstein's general theory of relativity was developed in November 1915~\cite{Einstein, Hilbert}, and within two months Karl Schwarzschild (working with one of the slightly earlier versions of the theory) had already solved the field equations that determine the exact spacetime geometry of a non-rotating ``point particle''~\cite{Schwarzschild1}.   It was relatively quickly realised,  via  Birkhoff's uniqueness theorem~\cite{Birkhoff, Jebsen, Deser, Ravndal}, that  the spacetime geometry in the vacuum region outside any localized spherically symmetric source is equivalent, up to a possible coordinate transformation, to a portion of the Schwarzschild geometry --- and so of direct physical interest to modelling the spacetime geometry surrounding and exterior to idealized non-rotating spherical stars and planets.
(In counterpoint, for modelling the \emph{interior} of a finite-size spherically symmetric source, Schwarzschild's ``constant density star'' is a useful first approximation~\cite{Schwarzschild2}. This is often referred to as Schwarzschild's ``interior'' solution, which is potentially confusing as it is an utterly distinct physical spacetime solving the Einstein equations in the presence of a specified distribution of matter.)

Considerably more slowly, only after intense debate was it realised that the ``inward'' analytic extension of Schwarzschild's ``exterior'' solution represents a non-rotating black hole, the endpoint of stellar collapse~\cite{Oppenheimer}. In the most common form (Schwarzschild coordinates, also known as curvature coordinates), which is not always the most useful form for understanding the physics, the Schwarzschild geometry is described by the line element
\begin{equation}
\d s^2 = - \left[1-{2m\over r}\right] \; \d t^2 + {\d r^2\over1-2m/r} + r^2 (\d\theta^2+\sin^2\theta\; \d\phi^2),
\end{equation}
where the parameter $m$ is the physical mass of the central object.

But astrophysically, we know that stars (and for that matter planets) rotate, and from the weak-field approximation to the Einstein equations we even know the approximate form of the metric at large distances from a stationary isolated body of mass $m$ and angular momentum $J$~\cite{Lense-Thirring, Pfister, Adler-Bazin-Schiffer, MTW, D'Inverno, Hartle, Carroll}. In suitable coordinates:
\begin{eqnarray}
\d s^2 &=& - \left[1-{2m\over r} + O\left({1\over r^2}\right)\right] \; \d t^2 
- \left[{4 J \sin^2\theta\over r} +  O\left({1\over r^2}\right)\right] \; \d\phi\;\d t
\nonumber
\\
&&
+ \left[1+ {2m\over r} + O\left({1\over r^2}\right)\right] \; \left[\d r^2
+ r^2 (\d\theta^2+\sin^2\theta\; \d\phi^2)\right].
\end{eqnarray}
This approximate metric is perfectly adequate for almost all solar system tests of general relativity, but there certainly are well-known astrophysical situations (\eg, neutron stars) for which this approximation is inadequate --- and so a ``strong field'' solution is physically called for. Furthermore, if a rotating star were to undergo gravitational collapse, then the resulting black hole would be expected to retain at least some fraction of its initial angular momentum --- thus suggesting on physical grounds that somehow there should be an extension of the Schwarzschild geometry to the situation where the central body carries angular momentum.

Physicists and mathematicians looked for such a solution for many years, and had almost given up hope, until the Kerr solution was discovered in 1963~\cite{Kerr} ---
some 48 years after the Einstein field equations were first developed. From the weak-field asymptotic result we can already see that angular momentum destroys spherical symmetry, and this 
lack of spherical symmetry makes the calculations \emph{much} more difficult.  It is not that the basic principles are all that different, but simply that the algebraic complexity of the computations is so high that relatively few physicists or mathematicians have the fortitude to carry them through to completion. 

Indeed it is easy to both derive and check the Schwarzschild solution by hand, but for the Kerr spacetime the situation is rather different.  
For  instance in Chandrasekhar's magnum opus on black holes~\cite{Chandrasekhar}, only part of which is devoted to the Kerr spacetime, he is moved to comment:
\begin{quotation}
``The treatment of the perturbations of the Kerr space-time in this chapter has been prolixius in its complexity. Perhaps, at a later time, the complexity will be unravelled by deeper insights. But meantime, the analysis has led us into a realm of the rococo: splendorous, joyful, and immensely ornate.''
\end{quotation}
More generally, Chandrasekhar also comments:
\begin{quote}
``The nature of developments  simply does not allow a presentation that can be followed in detail with modest effort: the reductions that are required to go from one step to another are often very elaborate and, on occasion, may require as many as ten, twenty, or even fifty pages.''
\end{quote}
Of course the Kerr spacetime was not constructed \emph{ex nihilo}. Some of Roy Kerr's early thoughts on this and related matters can be found in~\cite{Kerr1, Kerr2}, and over the years he has periodically revisited this theme~\cite{Kerr3, Kerr4, Kerr5, Kerr6, Kerr7, Kerr8, Kerr9, Kerr10}.

For practical and efficient computation in the Kerr spacetime many researchers will prefer to use general symbolic manipulation packages such as {\sf Maple}, {\sf Mathematica}, or more specialized packages such as {\sf GR-tensor}.  When used with \emph{care} and \emph{discretion}, symbolic manipulation tools can greatly aid physical understanding and insight.\footnote{For instance, the standard distribution of {\sf Maple} makes some unusual choices for its sign conventions. The signs of the Einstein tensor, Ricci tensor, and Ricci scalar (though \emph{not} the Riemann tensor and Weyl tensor) are opposite to what most physicists and mathematicians would expect.}

Because of the complexity of calculations involving the Kerr spacetime there is relatively little textbook coverage dedicated to this topic. An early discussion can be found in the textbook by Adler, Bazin, and Schiffer~\cite[1975 second edition]{Adler-Bazin-Schiffer}. The only dedicated single-topic textbook I know of is that by O'Neill~\cite{O'Neill}.  There are also comparatively brief discussions in the research monograph by Hawking and Ellis~\cite{Hawking-Ellis}, and the standard textbooks by Misner, Thorne, and Wheeler~\cite{MTW},  D'Inverno~\cite{D'Inverno}, Hartle~\cite{Hartle}, and Carroll~\cite{Carroll}.  One should particularly note the 60-page chapter appearing in the very recent 2006 textbook by Pleba\'nski and Krasi\'nski~\cite{Plebanski}.
An extensive and highly technical discussion of Kerr black holes is given in Chadrasekhar~\cite{Chandrasekhar}, while an exhaustive discussion of the class of spacetimes described by Kerr--Schild metrics is presented in the book ``Exact Solutions to Einstein's Field Equations''~\cite{exact}.  

To orient the reader I will now provide some general discussion, and explicitly present the line element for the Kerr spacetime in its most commonly used coordinate systems. (Of course the physics cannot depend on the coordinate system, but specific computations can sometimes be simplified by choosing an appropriate coordinate chart.)
  
\section{No Birkhoff theorem}

Physically, it must be emphasised that there is no Birkhoff theorem for rotating spacetimes --- it is  \emph{not} true that the spacetime geometry in the vacuum region outside a generic rotating star (or planet)
  is a part of the Kerr geometry. 
  The best result one can obtain is the much milder statement that outside a
  rotating star (or planet) the geometry asymptotically approaches the Kerr
  geometry.
  
  The basic problem is that in the Kerr geometry all the multipole
  moments are very closely related to each other --- whereas in real
  physical stars (or planets) the mass quadrupole, octopole, and higher moments of
  the mass distribution can in principle be independently specified.
  Of course from electromagnetism you will remember that higher
  $n$-pole fields fall off as $1/r^{2+n}$, so that far away from the
  object the lowest multipoles dominate --- it is in this \emph{asymptotic} sense that the 
  Kerr geometry is relevant for rotating stars or planets.
  
  On the other hand, if the star (or planet) gravitationally collapses --- then classically a black hole can be formed. In this case there \emph{are} a number of powerful uniqueness theorems which guarantee the direct physical relevance of the Kerr spacetime, but  as the unique exact solution corresponding to stationary rotating black holes,  (as opposed to merely being an asymptotic solution to the far field of rotating stars or planets).

\section{Kerr's original coordinates}

The very first version of the Kerr spacetime geometry to be explicitly written down in the literature was the line element~\cite{Kerr}
\begin{eqnarray}
\d s^2 &=& -\left[ 1 - {2mr\over r^2+a^2\cos^2\theta}\right]\; \left(\d u + a \sin^2\theta \; \d \phi\right)^2
\nonumber
\\
&&
+2 \left(\d u + a \sin^2\theta \; \d \phi\right) \; \left(\d r + a \sin^2\theta \; \d \phi\right)
\nonumber
\\
&&
+ (r^2+a^2\cos^2\theta)\; (\d\theta^2+\sin^2\theta\;\d\phi^2).
\label{E:K1}
\end{eqnarray}
The key features of this spacetime geometry are:
\begin{itemize}

\item Using symbolic manipulation software it is easy to verify that this manifold is Ricci flat, $R_{ab}=0$, and so satisfies the vacuum Einstein field equations. Verifying this by hand is at best tedious.

\item There are three off-diagonal terms in the metric --- which is one of the features that makes computations tedious.

\item By considering (for instance) the $g_{uu}$ component of the metric, 
it is clear that for $m\neq 0$ there is 
(at the very least) a coordinate singularity located at $r^2+a^2\,\cos^2\theta=0$, that is:
\begin{equation}
r=0;  \qquad \theta = \pi/2.
\end{equation}
We shall soon see that this is actually a curvature singularity.
In these particular coordinates there are no other obvious coordinate singularities. 

\item Since the line element is independent of both $u$ and $\phi$ we immediately deduce the existence of two Killing vectors. Ordering the coordinates as $(u,r,\theta,\phi)$ the two Killing vectors are
\begin{equation}
U^a = (1,0,0,0);  \qquad R^a=(0,0,0,1).
\end{equation}
Any constant-coefficient linear combination of these Killing vectors will again be a Killing vector.

\item
Setting $a\to0$ the line element reduces to
\begin{eqnarray}
\label{E:Sch-EF}
\d s^2 &\to& -\left[ 1 - {2m\over r}\right]\; \d u^2
+2\; \d u \; \d r 
+ r^2 \; (\d\theta^2+\sin^2\theta\;\d\phi^2),
\end{eqnarray}
which is the Schwarzschild geometry in the so-called  ``advanced Edding\-ton--Finkelstein coordinates''.  Based on this, by analogy the line element (\ref{E:K1}) is often called the advanced Eddington--Finkelstein form of the Kerr  spacetime. Furthermore since we know that $r=0$ is a curvature singularity in the Schwarzschild geometry, this strongly suggests (but does not yet prove) that the singularity in the Kerr spacetime at ($r=0$, $\theta=\pi/2$) is a curvature singularity.

\item
Setting $m\to 0$ the line element reduces to
\begin{eqnarray}
\label{E:M_EF}
\d s^2 \to \d s^2_0 &=& - \left(\d u + a \sin^2\theta \; \d \phi\right)^2
\nonumber
\\
&&
+2 \left(\d u + a \sin^2\theta \; \d \phi\right) \; \left(\d r + a \sin^2\theta \; \d \phi\right)
\nonumber
\\
&&
+ (r^2+a^2\cos^2\theta)\; (\d\theta^2+\sin^2\theta\;\d\phi^2),
\end{eqnarray}
which is actually (\emph{but certainly not obviously}) flat Minkowski spacetime in disguise. This is most easily seen by using symbolic manipulation software to verify that for this simplified line element the Riemann tensor is identically zero: $R_{abcd}\to0$.

\item For the general situation, $m\neq0\neq a$, all the non-zero components of the Riemann tensor contain at least one factor of $m$. 

\item Indeed, in a suitably chosen orthonormal basis the result can be shown to be even stronger: All the non-zero components of the Riemann tensor are then proportional to $m$:
\begin{equation}
R_{\hat a\hat b\hat c\hat d} \propto m.
\end{equation}
(This point will be discussed more fully below, in the section on the rational polynomial form of the Kerr metric. See also the discussion in~\cite{Plebanski}.) 

\item
Furthermore, the only nontrivial quadratic curvature invariant is
\begin{eqnarray}
R_{abcd} \; R^{abcd} &=&     C_{abcd} \; C^{abcd} \\
&=&
 {48m^2(r^2-a^2\cos^2\theta)\;[(r^2+a^2\cos^2\theta)^2-16r^2a^2\cos^2\theta)]\over (r^2+a^2\cos^2\theta)^6},
 \nonumber
\end{eqnarray}
guaranteeing that the singularity located at
\begin{equation}
r=0;  \qquad \theta = \pi/2,
\end{equation}
is actually a curvature singularity. (We would have strongly suspected this by considering the $a\to0$ case above.)

\item In terms of the $m=0$ line element we can put the line element into manifestly Kerr--Schild form by writing
\begin{equation}
\d s^2 = \d s_0^2 + {2mr\over r^2+a^2\cos^2\theta}\; \left(\d u + a \sin^2\theta \; \d \phi\right)^2,
\end{equation}
or the equivalent
\begin{equation}
g_{ab} = (g_0)_{ab} + {2mr\over r^2+a^2\cos^2\theta}\; \ell_a \; \ell_b,
\end{equation}
where we  define
\begin{equation}
(g_0)_{ab} = 
\left[
\begin{array}{cccc}
-1  & 1  & 0 &0  \\
1  &  0 & 0 & a \sin^2\theta  \\
0  & 0  & r^2+a^2\cos^2\theta & 0\\
0 & a\sin^2\theta & 0 &  (r^2+a^2) \sin^2\theta
\end{array}
\right],
\end{equation}
and
\begin{equation}
\ell_a = (1,0,0,a\sin^2\theta).
\end{equation}

\item
It is then easy to check that $\ell_a$ is a null vector, with respect to both $g_{ab}$ and $(g_0)_{ab}$, and that
\begin{equation}
g^{ab} = (g_0)^{ab} - {2mr\over r^2+a^2\cos^2\theta}\; \ell^a \; \ell^b,
\end{equation}
where
%
\begin{equation}
(g_0)^{ab} = {1\over r^2+a^2\cos^2\theta}
\left[
\begin{array}{cccc}
a^2\sin^2\theta  & r^2+a^2 &  
 0& -a
 \\
r^2+a^2  &  r^2+a^2   &
    0&  -a
    \\
0  &  0  & 1  & 0
\\
 -a &    -a & 
 0  &  (\sin^2\theta)^{-1}
\end{array}
\right],
\end{equation}
and
\begin{equation}
\ell^a = (0,1,0,0).
\end{equation}

\item
The determinant of the metric takes on a remarkably simple form
\begin{equation}
\det(g_{ab}) = - (r^2+a^2\;\cos^2\theta)^2 \sin^2\theta = \det([g_0]_{ab}),
\end{equation}
where the $m$ dependence has cancelled. (This is a side effect of the fact that the metric is of the Kerr--Schild form.)

\item 
At the curvature singularity we have
\begin{equation}
\left.\d s_0^2\right|_\mathrm{singularity} = - \d u^2 + a^2 \; \d \phi^2,
\end{equation}
showing that, in terms of the ``background'' geometry specified by the disguised Minkowski spacetime with metric $(g_0)_{ab}$, the curvature singularity is a ``ring''.
Of course in terms of the ``full''  geometry, specified by the physical metric $g_{ab}$, the intrinsic geometry of the curvature singularity is, unavoidably and by definition, singular.

\item 
The null vector field $\ell^a$ is an affinely parameterized null geodesic:
\begin{equation}
\ell^a \; \nabla_a \ell^b = 0.
\end{equation}

\item
More generally
\begin{equation}
\ell_{(a;b)} =  
\left[
\begin{array}{cccc}
0  & 0  & 0 & 0  \\
0  & 0  & 0 & 0  \\
0  & 0  & r & 0\\
0  & 0  & 0 & r\sin^2\theta  
\end{array}
\right]_{ab}
- {m(r^2-a^2\cos^2\theta)\over(r^2+a^2\cos^2\theta)^2} \; \ell_a\;\ell_b.
\end{equation}
(And it is easy to see that this automatically implies the null vector field $\ell^a$ is an affinely parameterized null geodesic.)

\item
The divergence of the null vector field $\ell^a$ is also particularly simple
\begin{equation}
\nabla_a \ell^a =  {2r\over r^2+a^2\;\cos^2\theta}.
\end{equation}

\item 
Furthermore, with the results we already have it is easy to calculate
\begin{equation}
\ell_{(a;b)} \; \ell^{(a;b)}  = \ell_{(a;b)} g^{bc} \ell_{(c;d)} g^{da} 
=  \ell_{(a;b)} [g_0]^{bc} \ell_{(c;d)} [g_0]^{da}
={2 r^2 \over(r^2+a^2\cos^2\theta)^2},
\end{equation}
whence
\begin{equation}
{\ell_{(a;b)} \; \ell^{(a;b)}\over2} - { (\nabla_a \ell^a )^2\over 4} = 0.
\end{equation}
This invariant condition implies that the null vector field $\ell^a$ is ``shear free''.

\item 
Define a one-form $\ell$ by
\begin{equation}
\ell = \ell_a \;\d x^a = \d u + a \sin^2\theta\;\d\phi,
\end{equation}
then
\begin{equation}
\d \ell = a \sin2\theta \;\d\theta\wedge\d\phi \neq 0,
\end{equation}
but implying
\begin{equation}
\d\ell \wedge \d\ell = 0.
\end{equation}
\item
Similarly
\begin{equation}
\ell\wedge\d \ell = a \sin2\theta \; \d u\wedge\d\theta\wedge\d\phi \neq 0,
\end{equation}
but in terms of the Hodge-star we have the invariant relation
\begin{equation}
*(\ell\wedge\d \ell) =     -{2a\cos\theta\over r^2+a^2\cos^2\theta} \; \ell,
\end{equation}
or in component notation
\begin{equation}
\epsilon^{abcd} (\ell_b\;\ell_{c,d}) =     -{2a\cos\theta\over r^2+a^2\cos^2\theta} \; \ell^a.
\end{equation}
This allows one to pick off the so-called ``twist'' as 
\begin{equation}
\omega = -{a\cos\theta\over r^2+a^2\cos^2\theta} 
\end{equation}

\end{itemize}
This list of properties is a quick, but certainly not exhaustive, survey of the key features of the spacetime that can be established by direct computation in this particular coordinate system.

\section{Kerr--Schild ``Cartesian'' coordinates}

The second version of the Kerr line element presented in the original article~\cite{Kerr}, also discussed in the early follow-up conference contribution~\cite{Kerr-Texas}, was defined in terms of ``Cartesian'' coordinates $(t,x,y,z)$:
\begin{eqnarray}
\label{E:K2}
\d s^2 &=&  - \d t^2 + \d x^2 + \d y^2 + \d z^2 
\\
&&+{2mr^3\over r^4+a^2 z^2} \Bigg[ \d t + {r(x \; \d x + y \; \d y ) \over a^2 + r^2} 
+{a(y \; \d x - x\;\d y) \over a^2 + r^2} + {z\over r} \d z \Bigg]^2,
\nonumber
\end{eqnarray}
subject to $r(x,y,z)$, which is now a dependent function not a coordinate, being implicitly determined by:
\begin{equation}
\label{E:r-KS}
x^2+y^2+z^2 = r^2 + a^2\left[1-{z^2\over r^2}\right].
\end{equation}
\begin{itemize}
\item 
The coordinate transformation used in going from (\ref{E:K1}) to  (\ref{E:K2}) is
\begin{equation}
\label{E:EF2KS}
 t = u - r; \qquad x+iy = (r-ia) \; e^{i\phi}\sin\theta; \qquad z=r\cos\theta.
\end{equation}
Sometimes it is more convenient to explicitly write
\begin{equation}
x= (r\cos\phi+a\sin\phi)\sin\theta = \sqrt{r^2+a^2} \;\sin\theta \cos[\phi-\arctan(a/r)];
\end{equation}
\begin{equation}
y= (r\sin\phi-a\cos\phi)\sin\theta = \sqrt{r^2+a^2} \; \sin\theta \sin[\phi-\arctan(a/r)];
\end{equation}
and so deduce
\begin{equation}
{x^2+y^2\over \sin^2\theta} - {z^2\over \cos^2\theta} = a^2,
\end{equation}
or the equivalent
\begin{equation}
{x^2+y^2\over r^2+a^2} + {z^2\over r^2} = 1.
\end{equation}

\item The $m\to0$ limit is now manifestly Minkowski space
\begin{eqnarray}
\label{E:M}
\d s^2 &\to& \d s_0^2 =  - \d t^2 + \d x^2 + \d y^2 + \d z^2.
\end{eqnarray}
Of course the coordinate transformation (\ref{E:EF2KS}) used in going from (\ref{E:K1}) to  (\ref{E:K2}) is also responsible for taking (\ref{E:M_EF}) to (\ref{E:M}).

\item  The $a\to0$ limit is 
\begin{eqnarray}
\d s^2 &\to&  - \d t^2 + \d x^2 + \d y^2 + \d z^2 
\\
&&+{2m\over r} \left[ \d t + {(x \; \d x + y \; \d y + z\; \d z)\over r} \right]^2,
\nonumber
\end{eqnarray}
now with 
\begin{equation}
r= \sqrt{x^2+y^2+z^2}.
\end{equation}
After a change of coordinates this can also be written as 
\begin{eqnarray}
\d s^2 &\to&  - \d t^2 + \d r^2 + r^2(\d\theta^2 + \sin^2\theta\;\d\phi^2) 
+{2m\over r} \left[ \d t + \d r \right]^2,
\end{eqnarray}
which is perhaps more readily recognized as the Schwarzschild spacetime. In fact if we set $u=t+r$ then 
we regain the Schwarzschild line element in advanced Eddington--Finkelstein coordinates; compare with equation (\ref{E:Sch-EF}).

\item The full $m\neq0\neq a$ metric is now manifestly of the Kerr--Schild  form
\begin{equation}
g_{ab} = \eta_{ab} + {2mr^3\over r^4+a^2 z^2}\;  \ell_a \; \ell_b,
\end{equation}
where now
\begin{equation}
\ell_a = \left(1, {rx+ay\over r^2+a^2}, {ry-ax\over r^2+a^2}, {z\over r} \right).
\end{equation}
Here $\ell_a$ is a null vector with respect to both $g_{ab}$ and $\eta_{ab}$ and
\begin{equation}
g^{ab} = \eta^{ab} - {2mr^3\over r^4+a^2 z^2}\;  \ell^a \; \ell^b,
\end{equation}
with
\begin{equation}
\ell^a = \left(-1, {rx+ay\over r^2+a^2}, {ry-ax\over r^2+a^2}, {z\over r} \right).
\end{equation}

\item
Define $R^2 = x^2+y^2+z^2$  then explicitly
\begin{equation}
r(x,y,z)  = \sqrt{R^2 - a^2 + \sqrt{(R^2-a^2)^2+4a^2z^2}\over2},
\end{equation}
where the positive root has been chosen so that $r\to R$ at large distances. Because of this relatively complicated expression for $r(x,y,z)$, direct evaluation of the Ricci tensor via symbolic manipulation packages is prohibitively expensive in terms of computer resources --- it is better to check the $R_{ab}=0$ in some other coordinate system, and then appeal to the known trivial transformation law for the zero tensor.

\item Consider the quadratic curvature invariant $R_{abcd} \; R^{abcd}= C_{abcd} \; C^{abcd}$, which we have already seen exhibits a curvature singularity at ($r=0$, $\theta=\pi/2$). In this new coordinate system the curvature singularity is located at
\begin{equation}
x^2 + y^2 = a^2; \qquad z=0.
\end{equation}
Again we recognize this as occurring on a ``ring'', now in the $(x,y,z)$ ``Cartesian'' background space.

\item In these coordinates the existence of a time-translation Killing vector, with components
\begin{equation}
K^a = (1,0,0,0)
\end{equation}
is obvious. Less obvious is what happens to the azimuthal (rotational) Killing vector, which now takes the form
\begin{equation}
 R^a=(0,0,y,-x).
\end{equation}

\item Note that in these coordinates
\begin{equation}
g^{tt} = -1 - {2m r^3\over r^4+a^2 z^2},
\end{equation}
and consequently
\begin{equation}
g^{ab} \; \nabla_a t \; \nabla_b t = -1 - {2m r^3\over r^4+a^2 z^2}.
\end{equation}
Thus $\nabla t$ is certainly a timelike vector in the region $r>0$, implying that this portion of the manifold is ``stably causal'', and that if one restricts attention to the region $r>0$ there is no possibility of forming closed timelike curves. However, if one chooses to work with the maximal analytic extension of the Kerr spacetime, then the region $r<0$ does make sense (at least mathematically), and certainly does contain closed timelike curves.  (See for instance the discussion in Hawking and Ellis~\cite{Hawking-Ellis}.)  Many (most?) relativists would argue that this $r<0$ portion of the maximally extended Kerr spacetime is purely of mathematical interest and  not physically relevant to astrophysical black holes.

\end{itemize}
So there is a quite definite trade-off in going to Cartesian coordinates --- some parts of the geometry are easier to understand, others are more obscure.

\section{Boyer--Lindquist coordinates}

Boyer--Lindquist coordinates are best motivated in two stages: 
First, consider a slightly different but completely equivalent form of the metric which follows from Kerr's original ``advanced Eddington--Finkelstein'' form (\ref{E:K1}) via the coordinate substitution
\begin{equation}
\label{E:BL-change-1}
u = t+r,
\end{equation}
in which case
\begin{eqnarray}
\label{E:K3}
\d s^2 &=& 
-\d t^2
\nonumber
\\
&&
 + \d r^2 + 2 a \sin^2\theta \;\d r \; \d \phi + (r^2+a^2\cos^2\theta) \; \d\theta^2 
+ (r^2+a^2) \sin^2\theta \;\d\phi^2
\nonumber
\\
&&
+{2mr\over r^2+a^2\cos^2\theta}\; \left(\d t + \d r + a \sin^2\theta \; \d \phi\right)^2.
\end{eqnarray}
Here the second line is again simply flat 3-space in disguise. An advantage of this coordinate system is that $t$ can naturally be thought of as a time coordinate --- at least at large distances near spatial infinity.
There are however still 3 off-diagonal terms in the metric so this is not yet any great advance on the original form  (\ref{E:K1}). One can easily consider the limits $m\to0$, $a\to0$, and the decomposition of this metric into Kerr--Schild form, but there are no real surprises.

Second, it is now extremely useful to perform a further \emph{$m$-dependent} coordinate transformation, which will put the line element  into Boyer--Lindquist form:
\begin{equation}
\label{E:BL-change-2}
t = t_{BL}+2m\int {r \;\d r\over r^2-2mr+a^2}; 
\qquad 
\phi = - \phi_{BL}-a \int {\d r\over r^2-2mr+a^2};
\end{equation}
\begin{equation}
\label{E:BL-change-3}
 r = r_{BL}; \qquad \theta=\theta_{BL}.
\end{equation}
Making the transformation, and dropping the $BL$ subscript, the Kerr line-element now takes the form:
\begin{eqnarray}
\d s^2 &=& - \left[ 1- {2mr\over r^2+a^2\cos^2\theta}\right] \d t^2 
- {4mra\sin^2\theta\over r^2+a^2\cos^2\theta}\; \d t\;\d\phi 
\\
&&
+ \left[{r^2+a^2\cos^2\theta\over  r^2-2mr+a^2}\right] \d r^2 
+ (r^2+a^2\cos^2\theta) \;\d\theta^2
\nonumber\\
&&
+ \left[r^2+a^2+ {2mr a^2 \sin^2\theta\over r^2+a^2\cos^2\theta}\right] \sin^2\theta\;\d\phi^2.
\nonumber
\end{eqnarray}
\begin{itemize}
\item
These Boyer--Lindquist coordinates are particularly useful in that they minimize the
number of off-diagonal components of the metric --- there is now only \emph{one} off-diagonal component. We shall subsequently see that this helps
particularly in analyzing the asymptotic behaviour, and in trying to understand the key difference between an ``event horizon'' and an ``ergosphere''.

\item
Another particularly useful feature is that the asymptotic ($r\to\infty$) behaviour in Boyer--Lindquist coordinates is
\begin{eqnarray}
\d s^2 &=& - \left[1-{2m\over r} + O\left({1\over r^3}\right)\right] \; \d t^2 
- \left[{4 ma  \sin^2\theta\over r} +  O\left({1\over r^3}\right)\right] \; \d\phi\;\d t
\nonumber
\\
&&
+ \left[1+{2m\over r}+ O\left({1\over r^2}\right)\right] \; \left[\d r^2
+ r^2 (\d\theta^2+\sin^2\theta\; \d\phi^2)\right].
\end{eqnarray}
From this we conclude that $m$ is indeed the mass and $J=ma$ is indeed the angular momentum.
\item 
If $a\to0$ the Boyer--Lindquist line element reproduces the Schwarzschild line element in standard Schwarzschild curvature coordinates.

\item
 If $m\to0$  Boyer--Lindquist line element reduces to
\begin{eqnarray}
\d s^2 &\to& 
-  \d t^2 
+ {r^2+ a^2\cos\theta^2\over r^2+a^2}\; \d r^2
\\
&&
+ (r^2+ a^2\cos\theta^2) \; \d\theta^2
+ (r^2+a^2)\; \sin^2\theta   \; \d\phi^2.
\nonumber
\end{eqnarray}
This is flat Minkowski space in  so-called ``oblate
spheroidal'' coordinates, and you can relate them to the usual Cartesian
coordinates of Euclidean 3-space by defining
\begin{eqnarray}
x &=& \sqrt{r^2+a^2} \; \sin\theta \;\cos\phi;
\\
y &=& \sqrt{r^2+a^2} \; \sin\theta \; \sin\phi;
\\
z &=& r \; \cos\theta.
\end{eqnarray}

\item One can re-write the Boyer--Lindquist line element as
\begin{eqnarray}
\d s^2 &=& 
-  \d t^2 
+ {r^2+ a^2\cos\theta^2\over r^2+a^2}\; \d r^2
\\
&&
+ (r^2+ a^2\cos\theta^2) \; \d\theta^2
+ (r^2+a^2)\; \sin^2\theta   \; \d\phi^2
\nonumber
\\
&&
+ {2m\over r} \left\{   { \left[\d t - a \sin^2\theta\; \d\phi\right]^2 \over  (1+ a^2\cos^2\theta/r^2)}
+ {(1+a^2\cos^2\theta/r^2) \; \d r^2\over1-2m/r+a^2/r^2} \right\}.
\nonumber
\end{eqnarray}
In view of the previous comment, this makes it clear that the Kerr
geometry is of the form (flat Minkowski space) + (distortion).  Note however that this is \emph{not} a Kerr--Schild decomposition for the Boyer--Lindquist form of the Kerr line element.

\item
Of course there will still be a Kerr--Schild decomposition of the metric
\begin{equation}
g_{ab} = (g_0)_{ab} + {2m\,r\over r^2+a^2\cos^2\theta}\; \ell_a \; \ell_b,
\end{equation}
but its form in Boyer--Lindquist coordinates is not so obvious. In these coordinates we have
\begin{equation}
\ell = \d t + {r^2+a^2\cos^2\theta\over r^2-2mr+a^2} \;\d r - a \sin^2\theta\;\d\phi,
\end{equation}
or the equivalent
\begin{equation}
\ell_a = \left(1, + {r^2+a^2\cos^2\theta\over r^2-2mr+a^2}, 0,  - a \sin^2\theta\right).
\end{equation}
Perhaps more surprisingly, at least at first glance, because of the $m$-dependent coordinate transformation used in going to Boyer--Lindquist coordinates the ``background'' metric $(g_0)_{ab}$ takes on the form:
\begin{equation}
\!\!\!
\left[
\begin{array}{cccc}
-1  & -{2mr\over r^2-2mr+a^2}  &  0 & 0\\
 -{2mr\over r^2-2mr+a^2}  &  {(r^2+a^2\cos^2\theta)(r^2-4mr+a^2)\over (r^2-2mr+a^2)^2} &  0 &  {2mar\sin^2\theta\over r^2-2mr+a^2}\\
0  &  0 &  r^2+a^2 \cos^2\theta & 0 \\
0  &  {2mar\sin^2\theta\over r^2-2mr+a^2} &  0  & (r^2+a^2)\sin^2\theta
\end{array}
\right].
\end{equation}
This is (arguably) the least transparent form of representing flat Min\-kow\-ski space that one is likely to encounter in any reasonably natural setting.
(That this is again flat Minkowski space in disguise can be verified by using symbolic manipulation packages to compute the Riemann tensor and checking that it is identically zero.) In short, while Boyer--Lindquist coordinates are well adapted to some purposes, they are very ill-adapted to probing the Kerr--Schild decomposition.

\item In these coordinates the time translation Killing vector is
\begin{equation}
K^a = (1,0,0,0),
\end{equation}
while the azimuthal Killing vector is
\begin{equation}
 R^a=(0,0,0,1).
\end{equation}

\item
 In Boyer--Lindquist coordinates, (because the $r$ and $\theta$ coordinates have not been modified, and because the Jacobian determinants arising from the coordinate transformations (\ref{E:BL-change-1}) and (\ref{E:BL-change-2})--(\ref{E:BL-change-3}) are both unity), the  determinant of the metric again takes the relatively simple $m$-independent form
\begin{equation}
\det(g_{ab}) = - \sin^2\theta \; (r^2+a^2\cos^2\theta)^2 = \det([g_0]_{ab}).
\end{equation}

\item In Boyer--Lindquist coordinates, (because the $r$ and $\theta$ coordinates have not been modified), the  invariant quantity $R_{abcd} \; R^{abcd}$ looks identical to that calculated for the line element (\ref{E:K1}).  Namely:
\begin{equation}
R_{abcd} \; R^{abcd} =  {48m^2(r^2-a^2\cos^2\theta)\;[(r^2+a^2\cos^2\theta)^2-16r^2a^2\cos^2\theta)]\over (r^2+a^2\cos^2\theta)^6}.
\end{equation}

\item On the axis of rotation ($\theta=0$, $\theta=\pi$), the Boyer--Lindquist line element reduces to:
\begin{eqnarray}
\left.\d s^2\right|_\mathrm{on-axis} &=& - \left[{1- {2m/ r}+a^2/r^2\over1+a^2/r^2}\right] \; \d t^2 
\nonumber\\
&&
+ \left[{1+a^2/r^2\over  1-2m/r+a^2/r^2}\right] \;\d r^2. 
\end{eqnarray}
This observation is useful in that it suggests that the on-axis geometry (and in particular the on-axis causal structure) is \emph{qualitatively} similar to that of the Reissner--Nordstr\"om geometry.

\item On the equator ($\theta=\pi/2$) one has
\begin{eqnarray}
\left.\d s^2\right|_\mathrm{equator} &=& - \left[ 1- {2m\over r}\right] \d t^2 
- {4ma\over r}\; \d t\;\d\phi 
\\
&&
+ {\d r^2\over  1-2m/r+a^2/r^2}
+ \left[r^2+a^2+ {2m a^2 \over r}\right] \;\d\phi^2.
\nonumber
\end{eqnarray}
Alternatively
\begin{eqnarray}
\left.\d s^2\right|_\mathrm{equator} &=& - \d t^2 + {\d r^2\over  1-2m/r+a^2/r^2} + (r^2+a^2)\;\d\phi^2
\nonumber
\\
&&
+{2m\over r}\; (\d t + a \;\d \phi)^2.
\end{eqnarray}
This geometry is still rather complicated and, in contrast to the on-axis geometry, has no ``simple'' analogue. (See, for example,~\cite{vortex}.)

\end{itemize}

\section{``Rational polynomial'' coordinates}

Starting from  Boyer--Lindquist coordinates, we can define a new coordinate by $\chi = \cos\theta$ so that $\chi \in [-1,1]$.
  Then $ \sin\theta = \sqrt{1-\chi^2}$ and
\begin{equation}
\d \chi = \sin\theta \; \d\theta 
\qquad \implies \qquad 
\d\theta = {\d \chi\over\sqrt{1-\chi^2}}
\end{equation}
In terms of these $(t,r,\chi,\phi)$ coordinates the Boyer--Lindquist
version of the Kerr spacetime becomes
\begin{eqnarray}
\d s^2 &=& 
- \left\{ 1 - {2m r\over r^2+ a^2 \chi^2} \right\} \; \d t^2 
-{4 amr (1-\chi^2)\over  r^2+ a^2 \chi^2} \; \d\phi\;\d t
\nonumber
\\
&&
+ {r^2+ a^2 \chi^2\over r^2-2mr+a^2}\; \d r^2
+ (r^2+ a^2 \chi^2) \; {\d \chi^2\over 1 -\chi^2}
\nonumber
\\
&&
+ (1-\chi^2)
\left\{ 
r^2 +a^2+ {2ma^2r(1-\chi^2)\over r^2+ a^2 \chi^2}
\right\}  \; \d\phi^2.
\end{eqnarray}
This version of the Kerr metric has the virtue that all the metric
components are now simple rational polynomials of the coordinates ---
the elimination of trigonometric functions makes computer-based symbolic computations much less resource intensive --- some calculations speed up by factors of 100 or more. 
\begin{itemize}
\item 
For instance, the quadratic curvature invariant
\begin{equation}
R^{abcd}\;R_{abcd} = {48(r^2-a^2\chi^2)[(r^2+a^2\chi^2)^2-16r^2a^2\chi^2]
\over(r^2+a^2\chi^2)^6},
\end{equation}
can (on a ``modern'' laptop, \emph{ca.} 2008) now be extracted in a fraction of a second --- as opposed to a minute or more when trigonometric functions are involved.\footnote{Other computer-based symbolic manipulation approaches to calculating the curvature invariants have also been explicitly discussed in~\cite{Lake-invariants}.}

\item
More generally, consider the (inverse) tetrad
\begin{equation}
e^A{}_a = 
\left[
\begin{array}{c}
 e^0{}_a\\
 e^1{}_a\\
 e^2{}_a\\
 e^3{}_a\\   
\end{array}
\right]
\end{equation}
specified by
\begin{equation}
\left[
\begin{array}{cccc}
\sqrt{r^2-2mr+a^2\over r^2+a^2\chi^2}  & 0  & 0 & -a(1-\chi^2) \sqrt{r^2-2mr+a^2\over r^2+a^2\chi^2}  \\
0  &  \sqrt{r^2+a^2\chi^2\over r^2-2mr+a^2} & 0 & 0  \\
0   & 0  & \sqrt{r^2+a^2\chi^2\over1-\chi^2} & 0\\
- a\sqrt{1-\chi^2\over r^2+a^2} & 0 & 0 &   \sqrt{1-\chi^2\over r^2+a^2} (r^2+a^2)
\end{array}
\right],
\end{equation}
which at worst contains square roots of rational polynomials.
Then it is easy to check
\begin{equation}
g_{ab} \; e^A{}_a \; e^B{}_b = \eta^{AB}; \qquad \eta_{AB} \; e^A{}_a \; e^B{}_b = g_{ab}.
\end{equation}

\item
Furthermore, the associated tetrad is
\begin{equation}
e_A{}^a = 
\left[
\begin{array}{c}
 e_0{}^a\\
 e_1{}^a\\
 e_2{}^a\\
 e_3{}^a\\   
\end{array}
\right]
\end{equation}
specified by
\begin{equation}
\!\!\!\!\!\!\!\!
\left[
\begin{array}{cccc}
{r^2+a^2\over\sqrt{(r^2-2mr+a^2)(r^2+a^2\chi^2)}}  & 0  & 0& {a\over\sqrt{(r^2-2mr+a^2)(r^2+a^2\chi^2)}} \\
0  &  \sqrt{r^2+a^2\chi^2\over r^2-2mr+a^2}  &0   &0 \\
0  & 0  & \sqrt{r^2+a^2\chi^2\over 1-\chi^2}    &0 \\
a\sqrt{1-\chi^2\over r^2+a^2}  & 0  & 0 & {1\over\sqrt{(1-\chi^2)(r^2+a^2)} }
\end{array}
\right].
\end{equation}

\item
In this orthonormal basis the non-zero components of the Riemann tensor take on only two distinct (and rather simple) values:
\begin{eqnarray}
R_{0101}&=&-2 R_{0202}  = -2 R_{0303} = 2 R_{1212} = 2 R_{1313} = - R_{2323} 
\nonumber\\
    &=&  -{2mr(r^2-3 a^2\chi^2)\over(r^2+a^2\chi^2)^3},
\end{eqnarray}
and
\begin{equation}
R_{0123}  = R_{0213} = - R_{0312} = {2m a \chi (3r^2-a^2\chi^2)\over(r^2+a^2\chi^2)^3}.
\end{equation}
(See the related discussion in~\cite{Plebanski}.)

\item
As promised these orthonormal components of the Riemann tensor are now linear in $m$.
Furthermore, note that the only place where \emph{any} of the orthonormal components becomes infinite is where $r=0$ and $\chi=0$ --- thus verifying that the ring singularity we identified by looking at the scalar $R_{abcs}\;R^{abcd}$ is indeed the whole story --- there are no other curvature singularities hiding elsewhere in the Kerr spacetime.  

\item On the equator, $\chi=0$, the only non-zero parts of the Riemann tensor are particularly simple
\begin{eqnarray}
R_{0101} &=&-2 R_{0202}  = -2 R_{0303} = 2 R_{1212} = 2 R_{1313} = - R_{2323} 
\nonumber
\\
&=&  -{2m\over r^3},
\end{eqnarray}
which are the same as would arise in the Schwarzschild geometry. 

\item On the axis of rotation, $\chi=\pm1$, we have
\begin{eqnarray}
R_{0101}&=&-2 R_{0202}  = -2 R_{0303} = 2 R_{1212} = 2 R_{1313} = - R_{2323} 
\nonumber\\
    &=&  -{2mr(r^2-3 a^2)\over(r^2+a^2)^3},
\end{eqnarray}
and
\begin{equation}
R_{0123}  = R_{0213} = - R_{0312} = \pm{2m a (3r^2-a^2)\over(r^2+a^2)^3}.
\end{equation}

\end{itemize}

\section{Doran coordinates}

The coordinates relatively recently introduced by Chris Doran (in 2000)~\cite{Doran} give yet another view on the Kerr spacetime.  This coordinate system was specifically developed to be as close as possible to the Painlev\'e--Gullstrand form of the Schwarzschild line element~\cite{Painleve, Gullstrand, analogue0}, a form that has become more popular over the last decade with continued developments in the ``analogue spacetime'' programme. 

Specifically, one takes the original Eddington--Finkelstein form of the Kerr line element (\ref{E:K1}) and performs the $m$-dependent substitutions
\begin{equation}
\d u = \d t + {\d r\over 1+ \sqrt{2mr/(r^2+a^2)}}; 
\end{equation}
\begin{equation}
\d\phi = \d\phi_\mathrm{Doran} + {a\;\d r\over r^2+a^2+\sqrt{2mr(r^2+a^2)}}.
\end{equation}
After dropping the subscript ``Doran'', in the new $(t,r,\theta,\phi)$ coordinates Doran's version of the Kerr line element takes the form~\cite{Doran}:
\begin{eqnarray}
\d s^2 &=& -\d t^2 + (r^2+a^2\cos^2\theta) \; \d\theta^2 + (r^2+a^2)\sin^2\theta\;\d\phi^2
\\
&& 
+\left[{r^2+a^2\cos^2\theta\over r^2+a^2}\right] \; 
\left\{ \d r + {\sqrt{2mr(r^2+a^2)}\over r^2+a^2\cos^2\theta}\; (\d t - a\sin^2\theta\; \d \phi) \right\}^2.
\nonumber
\end{eqnarray}
Key features of this line element are:
\begin{itemize}
\item 
As $a\to 0$ one obtains
\begin{equation}
\d s^2 \to -\d t^2 + \left\{ \d r + {\sqrt{2m\over r}}\; \d t  \right\}^2+ r^2(\d\theta^2 + \sin^2\theta\;\d\phi^2),
\end{equation}
which is simply Schwarzschild's geometry in Painlev\'e--Gullstrand form~\cite{Painleve, Gullstrand, analogue0}.

\item
As $m\to 0$ one obtains
\begin{eqnarray}
\d s^2 \to \d s_0^2 &=& -\d t^2 
+\left[{r^2+a^2\cos^2\theta\over r^2+a^2}\right] \; \d r^2
\\
&&
+ (r^2+a^2\cos^2\theta) \; \d\theta^2 + (r^2+a^2)\sin^2\theta\;\d\phi^2
,
\nonumber
\end{eqnarray}
which is flat Minkowski space in oblate spheroidal coordinates.

\item
Verifying that the line element is Ricci flat is best done with a symbolic manipulation package.

\item 
Again, since the $r$ and $\theta$ coordinates have not changed, the explicit formulae for $\det(g_{ab})$ and $R^{abcd}\;R_{abcd}$ are unaltered from those obtained in the Eddington--Finkelstein or Boyer--Lindquist coordinates.

\item
A nice feature of the Doran  form of the metric is that for the contravariant inverse metric
\begin{equation}
g^{tt} = -1.
\end{equation}
In the language of the ADM formalism, the Doran coordinates slice the Kerr spacetime in such a way that the ``lapse'' is everywhere unity. This can be phrased more invariantly as the statement
\begin{equation}
g^{ab} \; \nabla_a t \; \nabla_b t = -1,
\end{equation}
implying
\begin{equation}
g^{ab} \; \nabla_a t \; \nabla_c \nabla_b t = 0,
\end{equation}
whence
\begin{equation}
(\nabla_a t) \; g^{ab} \; \nabla_b ( \nabla_c t) = 0.
\end{equation}
That is, the vector field specified by 
\begin{equation}
W^a = \nabla^a t = g^{ab}\; \nabla_b t  = g^{ta}
\end{equation}
is an affinely parameterized timelike geodesic. Integral curves of the vector $W^a = \nabla^a t$  (in the Doran  coordinate system) are thus the closest analogue the Kerr spacetime has to the notion of a set of ``inertial frames'' that are initially rest at infinity, and are then permitted to free-fall towards the singularity.

\item 
Because of this unit lapse feature, it might at first seem that in Doran coordinates the Kerr spacetime is stably causal and that causal pathologies are thereby forbidden.  However the coordinate transformation leading to the Doran form of the metric breaks down in the region $r<0$ of the maximally extended Kerr spacetime. (The coordinate transformation becomes complex.) So there is in fact complete agreement between Doran and Kerr--Schild forms of the metric: The region $r>0$ is stably causal, and timelike curves confined to lie in the region $r>0$ cannot close back on themselves.

\end{itemize}

There is also a ``Cartesian'' form of Doran's line element in $(t,x,y,z)$ coordinates. Slightly modifying the presentation of~\cite{Doran}, the line element is  given by
\begin{equation}
\d s^2 = -\d t^2 + \d x^2 + \d y^2 + \d z^2 + 
\left(F^2 \; V_a \; V_b + F [V_a S_b + S_a V_b] \right) \; \d x^a \; \d x^b,
\end{equation}
where 
\begin{equation}
F = \sqrt{2mr\over r^2+ a^2}
\end{equation}
and where $r(x,y,z)$ is the same quantity as appeared in the Kerr--Schild coordinates, \emph{cf.} equation (\ref{E:r-KS}). Furthermore
\begin{equation}
V_a = \sqrt{r^2(r^2+a^2)\over r^2+a^2 z^2} \left( 1, {ay\over r^2+a^2}, {-ax\over r^2+a^2}, 0\right),
\end{equation}
and
\begin{equation}
S_a = \sqrt{r^2(r^2+a^2)\over r^4+ a^2 z^2} 
\left( 0, {r x\over r^2+a^2}, {r y\over r^2+a^2}, {z\over r}\right).
\end{equation}
It is then easy to check that $V$ and $S$ are orthonormal  with respect to the ``background'' Minkowski metric $\eta_{ab}$:
\begin{equation}
\eta^{ab} \; V_a V_b = -1;  \qquad \eta^{ab} \; V_a S_b = 0; \qquad \eta^{ab} \; S_a S_b = +1.
\end{equation}
These vectors are \emph{not} orthonormal with respect to the full metric $g_{ab}$ but are nevertheless useful for investigating principal null congruences~\cite{Doran}.

\section{Other coordinates?}

We have so far seen various common coordinate systems that have been used over the past 44 years to probe the Kerr geometry.
As an open-ended exercise, one could find [either via {\sf Google}, the {\sf arXiv}, the
  library, or original research] as many different coordinate systems
  for Kerr as  one could stomach. Which of these coordinate systems is the
  ``nicest''? This is by no means obvious.  It is conceivable (if unlikely) that other coordinate systems may be found in the future that might make computations in the Kerr spacetime significantly simpler.

%
\begin{figure}[ht]
 \begin{center}
 \input{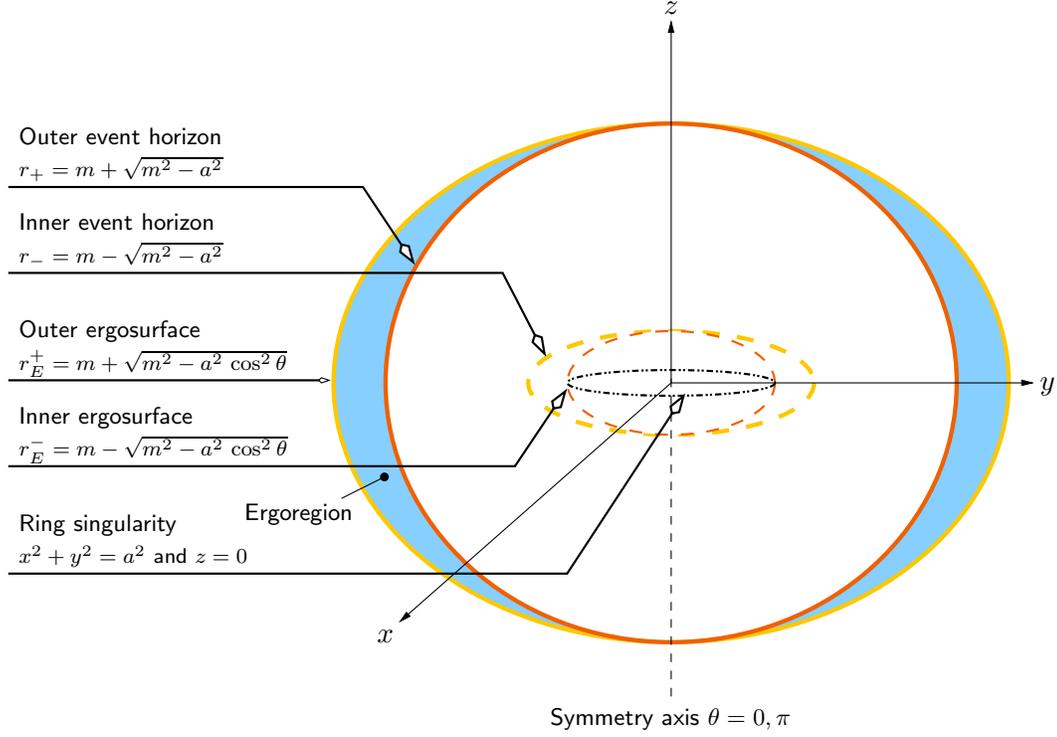}
 \caption[Rotating Kerr black hole.]  {\label{F:silke}
 Schematic location of the horizons, ergosurfaces, and curvature singularity in the Kerr spacetime.
 }
 \end{center}
\end{figure}
%

\section{Horizons}

To briefly survey the key properties of the horizons and ergospheres occurring in the Kerr spacetime, let us concentrate on using the   Boyer--Lindquist coordinates.
First, consider the components of the metric in these  coordinates.
The metric components have singularities when either
\begin{equation}
r^2 + a^2 \sin^2\theta=0, \qquad\hbox{that is,} \qquad r=0 \quad\hbox{and}\quad \theta=\pi/2,
\end{equation}
or
\begin{equation}
1-2m/r+a^2/r^2=0,  \qquad\hbox{that is,} \qquad r= r_\pm \equiv m\pm\sqrt{m^2-a^2}.
\end{equation}
The first of these possibilities corresponds to what we have already seen is a real physical curvature 
singularity, since $R_{abcd}\;R^{abcd}\to\infty$ there. In contrast
$R_{abcd}\;R^{abcd}$ remains finite at
$r_\pm$.

In fact the second option above ($r=r_\pm$) corresponds to \emph{all} orthonormal curvature components and \emph{all} curvature invariants being
finite --- it is a coordinate singularity but not a curvature
singularity. Furthermore as $a\to 0$ we have the smooth limit that
$r_\pm \to 2m$, the location of the horizon in the Schwarzschild
geometry. We will therefore tentatively identify $r_\pm$ as the
locations of the inner and outer horizons. (I shall restrict attention to the case $a<m$ to avoid having to deal with either the extremal case $a=m$ or the naked singularities that occur for $a>m$.)

Consider now the set of 3-dimensional surfaces formed by fixing the
$r$ coordinate to take on some specified fixed value $r=r_*$, and letting the
other three coordinates $(t,\theta,\phi)$ run over their respective ranges. On
these 3-dimensional surfaces the induced metric is found by formally
setting $\d r\to 0$ and $r\to r_*$ so that
\begin{eqnarray}
\d s^2_{\mathrm{3-surface}} &=& 
-  \d t^2 
+ (r_*^2+ a^2\cos\theta^2) \; \d\theta^2
+ (r_*^2+a^2)\; \sin^2\theta   \; \d\phi^2
\nonumber
\\
&&
+ {2m\over r_*}   { \left[\d t - a \sin^2\theta\; \d\phi\right]^2 \over  (1+ a^2\cos^2\theta/r_*^2)}.
\end{eqnarray}
Now we could calculate the determinant of this 3-metric directly.
Alternatively we could use the the block-diagonal properties of the metric in Boyer--Lindquist coordinates plus the fact that for the full (3+1)
dimensional metric we have already observed
\begin{equation}
\det(g_{ab}) = - \sin^2\theta \; (r^2+a^2\cos^2\theta)^2,
\end{equation}
to allow us to deduce that on the 3-surface $r=r_*$:
\begin{equation}
\det(g_{ij})_{\mathrm{3-surface}} = - \sin^2\theta \; (r_*^2+a^2\cos^2\theta) \; (r_*^2-2mr_*+a^2).
\end{equation}

Now for $r_*>r_+$ or $r_*<r_-$ the determinant
$\det(g_{ij})_{\mathrm{3-surface}}$ is negative, which is a
necessary condition for these 3-surfaces to be (2+1) dimensional [two
space plus one time dimension]. For $r_*\in(r_-,r_+)$ the determinant
$\det(g_{ij})_{\mathrm{3-surface}}$ is positive, indicating
the lack of any time dimension --- the label ``$t$'' is now misleading
and ``$t$'' actually denotes a spacelike direction. Specifically at what we shall soon see are 
the inner and outer horizons, $r_* = r_\pm$, the determinant is zero ---
indicating that the induced 3-metric
$[g_{ij}]_{\mathrm{3-surface}}$ is singular [in the sense of
being represented by a singular matrix, \emph{not} in the sense that there is a curvature singularity] everywhere on both of these
3-surfaces.

In particular, since $[g_{ij}]$ is a singular matrix, then at each
point in either of the 3-surface at $r=r_\pm$ there will be some  3-vector $L^i$ that lies in
the 3-surface $r=r_\pm$ such that
\begin{equation}
[g_{ij}] \; L^i = 0,
\end{equation}
which implies in particular
\begin{equation}
[g_{ij}] \; L^i \; L^j = 0.
\end{equation}
Now promote the 3-vector $L^i$ [which lives in 3-dimensional $(t,\theta,\phi)$
space] to a 4-vector by tacking on an extra coefficient that has value
zero:
\begin{equation}
L^i \to L^a = (L^t,0,L^\theta,L^\phi).
\end{equation}
Then in the (3+1) dimensional sense we have
\begin{equation}
g_{ab} \; L^a \;  L^b = 0 \qquad \hbox{($L^a$ only defined at $r=r_\pm$})
\end{equation}
That is: There is a set of curves, described by the vector
$L^a$, that lie precisely on the 3-surfaces $r=r_\pm$ and which do not
leave those 3-surfaces. Furthermore on the surfaces $r=r_\pm$ these curves are null curves and $L^a$ is a null vector.  Note that these null vector fields $L^a$ are defined only on the inner an outer horizons $r=r_\pm$ and that they are quite distinct from the null vector field $\ell^a$ occurring in the Kerr--Schild decomposition of the metric, that vector field being defined throughout the entire spacetime.

Physically, the vector fields $L^a$ correspond to photon ``orbits'' that skim along the
surface if the inner and outer horizons without either falling in or
escaping to infinity.
You should then be able to easily convince yourself that the outer
horizon is an ``event horizon'' [``absolute horizon''] in the sense of
being the boundary of the region from which null curves do not escape
to infinity, and we shall often concentrate discussion on the outer horizon $r_+$.

Indeed if we define quantities $\Omega_\pm$  by
\begin{equation}
\Omega_\pm = {a\over2m r_\pm } = {a\over r_\pm^2+a^2},
\end{equation}
and now define
\begin{equation}
L_\pm^a =  (L_\pm^t,0,0,L_\pm^\phi) = (1,0,0,\Omega_\pm),
\end{equation}
then it is easy to check that 
\begin{equation}
g_{ab}(r_\pm) \; L_\pm^a \;  L_\pm^b = 0
\end{equation}
at the 3 surfaces $r=r_\pm$ respectively. 
That is, the spacetime curves
\begin{equation}
X(t) = \left(t,r(t),\theta(t),\phi(r)\right) = 
\left(t, \;r_\pm, \; \theta_0, \; \phi_0 + \Omega_\pm\; t\right)
\end{equation}
are an explicit set of null curves (so they could represent photon
trajectories) that skim along the 3-surfaces $r=r_\pm$, the outer and inner horizons.  

In fact $r_+$ is the ``event
horizon'' of the Kerr black hole, and $\Omega_+$ is a constant over the 3-surface $r=r_+$. This is a consequence  of the ``rigidity theorems'', and the 
quantity $\Omega_+$ is referred to as the ``angular velocity'' of the
event horizon. The event horizon ``rotates'' \emph{as though} it were a solid body.
(Somewhat counter-intuitively, the inner horizon also rotates as though it were a solid body, but with a different angular velocity $\Omega_-$.)

Because the inner and outer horizons are specified by the simple condition
$r=r_\pm$, you might be tempted to deduce that the horizons are
``spherical''. Disabuse yourself of this notion. We have already
looked at the induced geometry of the 3-surface $r=r_\pm$ and found the
induced metric to be described by a singular matrix. Now let's
additionally throw away the $t$ direction, and look at the 2-surface
$r=r_\pm$, $t=0$, which is a 2-dimensional ``constant time'' slice
through the horizon. Following our previous discussion for 3-surfaces
the intrinsic 2-geometry of this slice is described by~\cite{Smarr}
\begin{eqnarray}
\d s^2_{\mathrm{2-surface}} &=& 
(r_\pm^2+ a^2\cos^2\theta) \; \d\theta^2
+ \left(4m^2 r_\pm^2\over  r_\pm^2+ a^2\cos^2\theta  \right) \; \sin^2\theta   \; \d\phi^2,
\nonumber
\\
&&
\end{eqnarray}
or equivalently
\begin{eqnarray}
\d s^2_{\mathrm{2-surface}} &=& 
(r_\pm^2+ a^2\cos^2\theta) \; \d\theta^2
+ \left( [r_\pm^2+a^2]^2 \over  r_\pm^2+ a^2\cos^2\theta  \right) \; \sin^2\theta   \; \d\phi^2.
\nonumber
\\
&&
\end{eqnarray}
So while the horizons are \emph{topologically} spherical, they are  emphatically not
\emph{geometrically} spherical. In fact the area of the horizons is~\cite{Smarr}
\begin{equation}
A_H^\pm = 4\pi( r_\pm^2+a^2) = 8\pi(m^2\pm\sqrt{m^4-m^2 a^2}) =  8\pi(m^2\pm\sqrt{m^4-L^2}).
\end{equation}
Furthermore, the Ricci scalar is~\cite{Smarr}
\begin{equation}
R_{\mathrm{2-surface}} = 
{2(r_\pm^2+a^2)\;(r_\pm^2-3a^2\cos^2\theta)\over(r_\pm^2+a^2\cos^2\theta)^3}.
\end{equation}
At the equator
\begin{equation}
R_{\mathrm{2-surface}} \to
{2(r_\pm^2+a^2)\over r_\pm^4} > 0,
\end{equation}
but at the poles
\begin{equation}
R_{\mathrm{2-surface}} \to
{2(r_\pm^2+a^2)\;(r_\pm^2-3a^2)\over(r_\pm^2+a^2)^3}.
\end{equation}
So the intrinsic curvature of the outer horizon (the event horizon) can actually be \emph{negative} near the axis  of rotation if $3a^2>r_+^2$, corresponding to $a>(\sqrt{3}/2)\; m$ --- and this happens \emph{before} one reaches the extremal case $a=m$ where the event horizon vanishes and a naked singularity forms. 

The Ricci 2-scalar of the outer horizon  is negative for 
\begin{equation}
\theta< \cos^{-1}\left({r_+\over\sqrt{3}a}\right),
\qquad\hbox{or}\qquad
\theta>\pi -  \cos^{-1}\left({r_+\over\sqrt{3}a}\right),
\end{equation}
which only has a non-vacuous range if  $a>(\sqrt{3}/2)\; m$. In contrast, on the inner horizon the Ricci 2-scalar goes negative for 
\begin{equation}
\theta< \cos^{-1}\left({r_-\over\sqrt{3}a}\right),
\qquad\hbox{or}\qquad
\theta>\pi -  \cos^{-1}\left({r_-\over\sqrt{3}a}\right),
\end{equation}
which always includes the region immediately surrounding the axis of rotation. In fact for slowly rotating Kerr spacetimes, $a\ll m$, the Ricci scalar goes negative on the inner horizon for the rather large range
\begin{equation}
\theta \lesssim \pi/2 - {1\over2\sqrt3} {a\over m}; \qquad 
\theta \gtrsim \pi/2 - {1\over2\sqrt3} {a\over m}.
\end{equation}

In contrast to the intrinsic geometry of the horizon, in terms of the Cartesian coordinates of the Kerr--Schild line element the location 
of the horizon is given by
\begin{equation}
x^2+y^2 + \left({r_\pm^2+a^2\over r_\pm^2}\right) z^2 = r_\pm^2+a^2, 
\end{equation}
which implies
\begin{equation}
x^2+y^2 + \left({2m\over r_\pm}\right) z^2 = 2 m r_\pm. 
\end{equation}
So in terms of the background geometry $(g_0)_{ab}$ the event horizons are a much simpler pair of oblate ellipsoids.
But like the statement that the singularity is a ``ring'', this is not a statement about the intrinsic geometry --- it is instead a statement about the mathematically convenient but fictitious flat Minkowski space that is so useful in analyzing the Kerr--Schild form of the Kerr spacetime. The semi-major axes of the oblate ellipsoids are
\begin{equation}
S_x = S_y = \sqrt{2m r_\pm} \leq 2 m;
\end{equation}
and
\begin{equation}
S_z = r_\pm \leq \sqrt{2m r_\pm} \leq 2 m.
\end{equation}
The eccentricity is
\begin{equation}
e_\pm =\sqrt{1-{S_z^2\over S_x^2}}=  \sqrt{1-{r_\pm\over2m}}.
\end{equation}
The volume interior to the horizons, with respect to the background Minkowski metric, is easily calculated to be
\begin{equation}
V_0^\pm = {4\pi\over3} (2m) r_\pm^2.
\end{equation}
The surface area of the horizons, with respect to the background Minkowski metric,  can then be calculated using standard formulae due to Lagrange:
\begin{equation}
A_0^\pm = 2 \pi (2m) r_\pm 
\left[ 1 + {1-e_\pm^2\over2e_\pm}\ln\left({1+e_\pm\over1-e_\pm}\right)\right].
\end{equation}
This however is not the intrinsic surface area, and is not the quantity relevant for the second law of black hole mechanics --- it is instead an illustration and warning of the fact that while the background Minkowski metric is often an aid to visualization, this background should not be taken seriously as an intrinsic part of the physics.

\section{Ergospheres}

There is a new concept for rotating black holes, the ``ergosphere'',
that does not arise for non-rotating black holes.  Suppose we have a
rocket ship and turn on its engines, and move so as to try to ``stand
still'' at a fixed point in the coordinate system --- that is, suppose
we try to follow the world line:
\begin{equation}
X(t) = \left(t,r(t),\theta(t),\phi(r)\right) = \left(t, r_0, \theta_0, \phi_0\right).
\end{equation}
Are there locations in the spacetime for which it is impossible to
``stand still'' (in this coordinate dependent sense)? Now the tangent vector to the world line of an observer who is
``standing still'' is
\begin{equation}
T^a = {\d X^a(t)\over \d t} = (1,0,0,0),
\end{equation}
and a necessary condition for a physical observer to be standing still
is that his 4-trajectory should be timelike. That is, we need
\begin{equation}
g(T,T) < 0.
\end{equation}
But
\begin{equation}
g(T,T) = g_{ab} \; T^a\; \; T^b = g_{tt},
\end{equation}
so in the specific case of the Kerr geometry (in Boyer--Lindquist coordinates) 
\begin{equation}
g(T,T) = -\left(1-{2mr\over r^2+a^2\cos^2\theta} \right).
\end{equation}
But
the RHS becomes positive once
\begin{equation}
r^2-2mr+a^2\cos^2\theta < 0.
\end{equation}
That is,  defining
\begin{equation}
r_E^\pm(m,a,\theta) \equiv m \pm \sqrt{m^2-a^2\cos^2\theta},
\end{equation}
the RHS becomes positive once
\begin{equation}
r_E^-(m,a,\theta) < r < r_E^+(m,a,\theta).
\end{equation} 
The surfaces $r= r_E^\pm(m,a,\theta)$, between which it is impossible to
stand still, are known as the ``stationary limit'' surfaces.

Compare this with the location of the event horizons
\begin{equation}
r_\pm \equiv m \pm \sqrt{m^2-a^2}.
\end{equation}
We see that 
\begin{equation}
r_E^+(m,a,\theta) \geq r_+ \geq r_- \geq r_E^-(m,a,\theta).
\end{equation}
In fact $r_E^+(m,a,\theta) \geq r_+$ with equality only at
$\theta=0$ and $\theta=\pi$ (corresponding to the axis of rotation). Similarly $r_E^-(m,a,\theta) \leq r_-$ with equality only at the axis of rotation. (This inner stationary limit surface
touches the inner horizon on the axis of rotation, but then plunges
down to the curvature singularity $r=0$ at the equator, $\theta=\pi/2$.)

Restricting attention to the ``outer'' region:  There is a region between the outer stationary limit surface and
the outer event horizon in which it is impossible to ``stand still'',
but it is still possible to escape to infinity. This region is known
as the ``ergosphere''.
Note that as rotation is switched off, $a\to0$, the stationary limit
surface moves to lie on top of the event horizon and the ergosphere
disappears.
(Sometimes one sees authors refer to the stationary limit surface as the
  ``ergosurface'', and to refer to the ergosphere as the ``ergoregion''.  Some authors furthermore use the word ``ergoregion'' to refer to the entire region between the stationary limit surfaces --- including the black hole region located between the horizons.)

\begin{itemize}
\item  By setting $r= r_E^\pm(m,a,\theta)$ and then replacing 
\begin{equation}
\d r \to \d r_E^\pm = \left({\d r_E^\pm\over\d\theta}\right) \; \d\theta = 
\pm {a^2 \cos\theta \sin\theta \over\sqrt{m^2-a^2\cos^2\theta}}\;\d\theta
\end{equation}
one can find the intrinsic induced 3-geometry on the ergosurface.  One form of the result is
\begin{eqnarray}
\d s^2_\mathrm{3-geometry} &=& 
- {2a\sin^2\theta}\; \d t\;\d\phi 
\\
&&
+ {2 m^3 r_E^\pm  \over m^2 - a^2 \cos^2\theta} \;\d\theta^2
+ 2  \left[ m r_E^\pm +a^2 \sin^2\theta\right] \sin^2\theta\;\d\phi^2.
\nonumber
\end{eqnarray}
This result is actually quite horrid, since if we wish to be explicit
\begin{eqnarray}
\d s^2_\mathrm{3-geometry} &=& 
- {2a\sin^2\theta}\; \d t\;\d\phi 
\\
&&
+ {2 m^3 (m\pm\sqrt{m^2-a^2\cos^2\theta})  \over m^2 - a^2 \cos^2\theta} \;\d\theta^2
\nonumber\\
&&
+ 2  \left[ m (m\pm\sqrt{m^2-a^2\cos^2\theta})  +a^2 \sin^2\theta\right] \sin^2\theta\;\d\phi^2.
\nonumber
\end{eqnarray}

\item  By now additionally setting  $\d t=0$ one can find the intrinsic induced 2-geometry on an instantaneous constant time-slice of the ergosurface~\cite{Lake}. One obtains
\begin{eqnarray}
\d s^2_\mathrm{2-geometry} &=& 
+ {2 m^3 (m\pm\sqrt{m^2-a^2\cos^2\theta})  \over m^2 - a^2 \cos^2\theta} \;\d\theta^2
\nonumber\\
&&
+ 2  \left[ m (m\pm\sqrt{m^2-a^2\cos^2\theta})  +a^2 \sin^2\theta\right] \sin^2\theta\;\d\phi^2.
\nonumber
\end{eqnarray}

\item
The intrinsic area of the ergosurface can then be evaluated as an explicit expression involving two incomplete  Elliptic integrals~\cite{Lake}. A more tractable approximate expression for the area of the outer ergosurface is~\cite{Lake}
\begin{equation}
A_E^+ = 4\pi\left[ (2m)^2 + a^2 + {3 a^4\over20 m^2}  + {33a^6\over 280m^4} +{191a^8\over 2880m^6} +  \mathcal{O}\left({a^{10}\over m^8}\right)  \right].
\end{equation}

\item
The 2-dimensional Ricci scalar is easily calculated but is again quite horrid~\cite{Lake}. A tractable approximation is
\begin{eqnarray}
R_E^+ &=& {1\over 2m^2} \left[ 1 + {3(1-6\cos^2\theta)a^2\over4m^2} 
- {3(1-5\cos^2\theta+3\cos^4\theta)a^4\over 8 m^4} \right.
\nonumber\\
&&
\qquad\qquad\qquad
\left. + \mathcal{O}\left({a^6\over m^6}\right) \right].
\end{eqnarray}
On the equator of the outer ergosurface there is a simple exact result that
\begin{equation}
R_E^+ (\theta=\pi/2) = {4m^2+5a^2\over4m^2(2m^2+a^2)}.
\end{equation}
At the poles the Ricci scalar has a delta function contribution coming from conical singularities at the north and south poles of the ergosurface. Near the north pole the metric takes the form
\begin{equation}
\label{E:ergo-pole}
\d s^2 =  K \left[ \d \theta^2 + \left(1-{a^2\over m^2}\right) \theta^2 \; \d\phi^2 
+ \mathcal{O}(\theta^4) \right],
\end{equation}
where $K$ is an irrelevant constant.

\item 
An isometric embedding of the outer ergosurface and a portion of the outer horizon in Euclidean 3-space is presented in figure~\ref{F:kayll}. Note the conical singularities at the north and south poles.

%
\begin{figure}[h!]
 \begin{center}
\includegraphics[width=10cm]{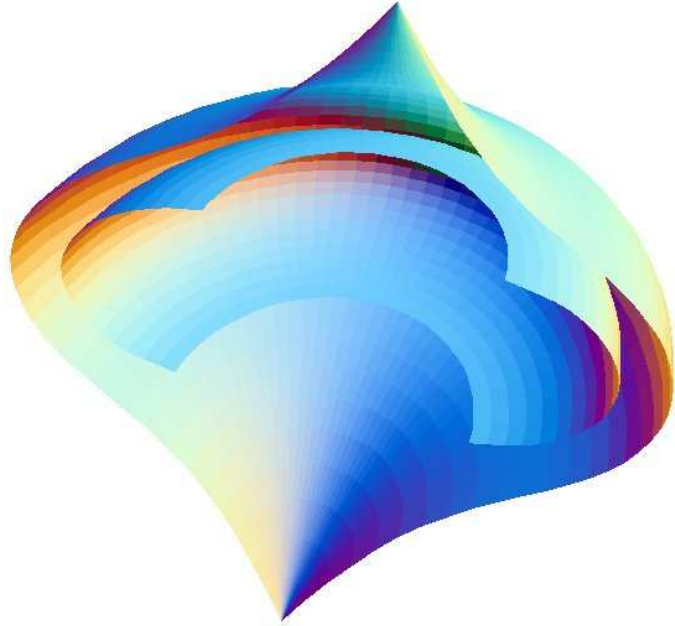}
 \caption[Rotating Kerr black hole.]  {\label{F:kayll}
Isometric embeddings in Euclidean space of the outer ergosurface and a portion of the outer horizon. For these images $a = 0.90\; m$. A cut to show the (partial) outer horizon, which is not isometrically embeddable around the poles, is shown. The polar radius of the ergosurface diverges for $a\to m$ as the conical singularities at the poles become more pronounced.  
 }
 \end{center}
\end{figure}
%

\item
The same techniques applied to the inner ergosurface yield
\begin{equation}
A_E^- = 4\pi\left[ {2\sqrt2-1\over3}  a^2 + {12\sqrt2-13 a^4\over20 m^2}  + {292\sqrt2-283 a^6\over 280m^4} 
+  \mathcal{O}\left({a^{8}\over m^6}\right)  
\right],
\end{equation}
and
\begin{eqnarray}
R_E^- &=& -{2(5-2\cos^2\theta)\over(2-\cos^2\theta)^2 a^2} 
-{76-198\cos^2\theta+128\cos^4\theta-25\cos^6\theta
\over2(8-12\cos^2\theta+6\cos^4\theta-\cos^6\theta) m^2} 
\nonumber\\
&&
\qquad\qquad\qquad
+ \mathcal{O}\left({a^2\over m^4}\right) .
\end{eqnarray}

\item
At the equator, the inner ergosurface touches the physical singularity. In terms of the intrinsic geometry of the inner ergosurface this shows up as the 2-dimensional Ricci scalar becoming infinite. The metric takes the form
\begin{equation}
\d s^2 = a^2(\theta-\pi/2)^2\d\theta^2 + a^2 \{2-3 (\theta-\pi/2)^2\}\;\d\phi^2 + \mathcal{O}[(\theta-\pi/2)^4],
\end{equation}
and after a change of variables can be written
\begin{equation}
\d s^2 = \d h^2 + \{2a^2-6 a |h|\}\;\d\phi^2 + \mathcal{O}[h^2].
\end{equation}
At the poles the inner ergosurface has conical singularities of the same type as the outer ergosurface. See equation (\ref{E:ergo-pole}).

\item   Working now in Kerr--Schild Cartesian coordinates, since the ergosurface is defined by the coordinate condition $g_{tt}=0$, we see that  this occurs at
\begin{equation}
r^4 - 2m r^3 + a^2 z^2 = 0.
\end{equation}
But recall that $r(x,y,z)$ is a rather complicated function of the Cartesian coordinates, so even in the ``background'' geometry the ergosurface is quite tricky to analyze. In fact it is better to describe the ergosurface parametrically by observing
\begin{equation}
\sqrt{x_E^2+y_E^2} = \sqrt{r_E(\theta)^2+ a^2} \;\sin\theta; \qquad z_E = r_E(\theta)\; \cos\theta.
\end{equation}
That is
\begin{equation}
\sqrt{x_E^2+y_E^2} = \left( \left[m\pm\sqrt{m^2-a^2\cos^2\theta}\right]^2+ a^2\right) \;\sin\theta; 
\end{equation}
\begin{equation}
z_E = (m\pm\sqrt{m^2-a^2\cos^2\theta}) \;\cos\theta.
\end{equation}
So again we see that the ergosurfaces are quite tricky to work with. See figure~\ref{F:cartesian} for a polar slice through the Kerr spacetime in these Cartesian coordinates.

%
\begin{figure}[h!]
 \begin{center}
 \includegraphics[width=11cm]{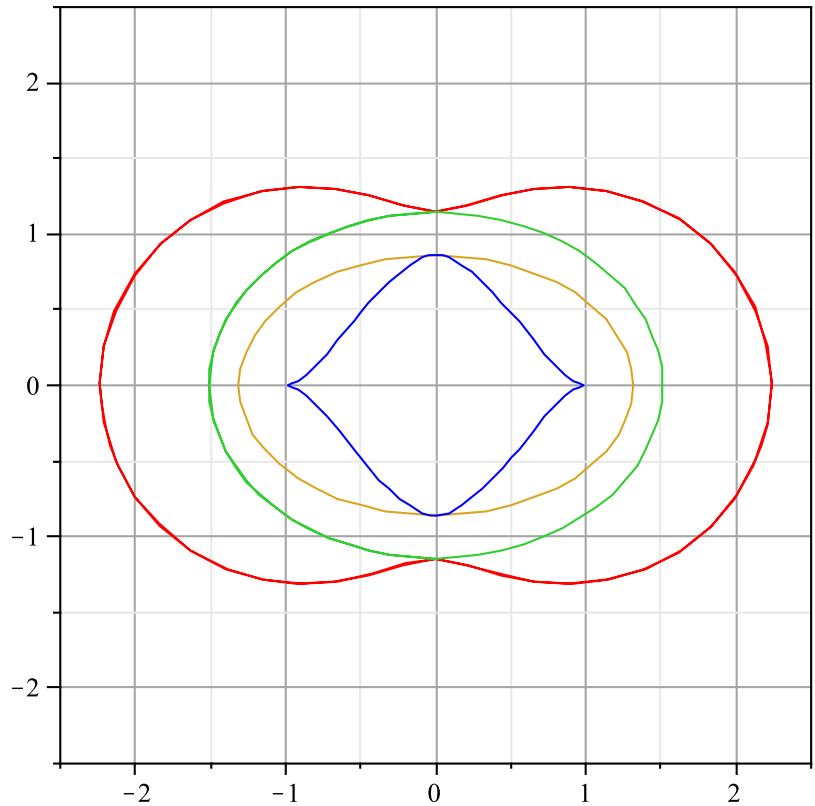}
 \caption[Polar slice through a Kerr black hole.]  {\label{F:cartesian}
Polar slice through the Kerr spacetime in Cartesian Kerr--Schild coordinates. 
Location of the  horizons, ergosurfaces, and curvature singularity is shown for $a=0.99\; m$ and $m=1$. Note that the inner and outer horizons are ellipses in these coordinates, while the inner and outer ergosurfaces are more complicated. The curvature singularity lies at the kink in the inner ergosurface.
 }
 \end{center}
\end{figure}
%

\end{itemize}

Putting these particular mathematical issues aside:  Physically and astrophysically, it is extremely important to realise that (assuming validity of the Einstein equations, which is certainly extremely reasonable given their current level of experimental and observational support) you should trust in the existence of the ergosurface and event horizon (that is, the outer ergosurface and outer horizon), and the region immediately below the event horizon.
  
 However you should \emph{not} physically trust in the inner horizon or the inner ergosurface. Although they are certainly there as mathematical solutions of the exact vacuum Einstein equations, there are good physics reasons to  suspect that the region at and inside the inner horizon, which can be shown to be a Cauchy horizon, is grossly unstable --- even classically --- and unlikely to form in any real
  astrophysical collapse.  
  
Aside from issues of stability, note that although the causal pathologies [closed timelike curves] in the Kerr spacetime have their genesis in the maximally extended $r<0$ region, the effects of these causal pathologies can reach out into part of the $r>0$ region, in fact out to the inner horizon at $r=r_-$~\cite{Hawking-Ellis} --- so the inner horizon is also a chronology horizon for the maximally extended Kerr spacetime. Just what does go on deep inside a classical or semiclassical black hole formed in real astrophysical collapse is still being debated --- see for instance the literature regarding ``mass inflation'' for some ideas~\cite{mass-inflation}. For astrophysical purposes it is certainly safe to discard the $r<0$ region, and almost all relativists would agree that it is safe to discard the entire region inside the inner horizon $r< r_-$.

\section{Killing vectors}

By considering the Killing vectors of the Kerr spacetime it is possible to develop more invariant characterizations of the ergosurfaces and horizons.
The two obvious Killing vectors are the time translation Killing vector $K^a$ and the azimuthal Killing vector $R^a$; in addition any constant coefficient linear combination $a \,K^a + b \,R^a$ is also a Killing vector --- and this exhausts the set of all Killing vectors of the Kerr spacetime.

The  time translation Killing vector $K^a$ is singled out as being the unique Killing vector that approaches  a unit timelike vector at spatial infinity. The vanishing of the norm
\begin{equation}
g_{ab} \; K^a \; K^b = 0
\end{equation}
is an invariant characterization of the ergosurfaces --- the ergosurfaces are where the time translation Killing vector becomes null.

Similarly, the horizon is a null 3-surface (since it is defined by a set of photon trajectories) that is further characterized by the fact  that it is invariant (since the photons neither fall into no escape from the black hole). This implies that there is \emph{some} Killing vector $a\, K^a + b\, R^a$ that becomes null on the horizon (a result that holds in any stationary spacetime, and leads to the ``rigidity theorem'').  Without loss of generality we can set $a\to 1$ and define $b\to \Omega_H$ so that the horizon is invariantly characterized by the condition
\begin{equation}
g_{ab} \; (K^a+\Omega_H\; R^a) \; (K^b+\Omega_H\;R^b) = 0.
\end{equation}
It is then a matter of computation to verify that this current definition of $\Omega_H$ coincides with that in the previous discussion.

\section{Comments}

There are a (large) number of things I have not discussed in this brief introduction, but
I have to cut things off somewhere. In particular, you can go to several textbooks, the original literature, and the rest of this book to see discussions of:
\begin{itemize}
\item The Kerr--Newman geometry (a charged rotating black hole).
\item Carter--Penrose diagrams of causal structure.
\item Maximal analytic extensions of the Kerr spacetime.
\item Achronal regions and chronology horizons  (time travel) in the idealized Kerr spacetime.
\item Extremal and naked Kerr spacetimes; cosmic censorship.
\item Particle and photon orbits in the Kerr black hole.
\item Geodesic completeness.
\item Physically reasonable sources and possible ``interior solutions'' for the Kerr spacetime.
\item Penrose process --- energy extraction from a rotating black hole.
\item Black hole uniqueness theorems.
\item Singularity theorems guaranteeing the formation of black holes in
  certain circumstances.
\item The classic laws of black hole mechanics; area increase theorem.
\item Black hole thermodynamics (Hawking temperature and Bekenstein entropy).
\end{itemize}
These and other issues continue to provoke considerable ongoing research --- and I hope this brief introduction will serve to orient the reader, whet one's appetite, and provoke interest in the rest of this book.

\section*{Acknowledgments}

Supported by the Marsden Fund administered by the Royal Society of New Zealand.
Figure \ref{F:silke} courtesy of Silke Weinfurtner. Figure \ref{F:kayll} courtesy of Kayll Lake. Additionally, I wish to acknowledge useful comments from Silke Weinfurtner, Kayll Lake, Bartolome Alles, and Roy Kerr.





\begin{thebibliography}{69}

\bibitem{Kerr}
Roy Kerr, 
``Gravitational field of a spinning mass as an example of algebraically special metrics'',
 Physical Review Letters {\bf  11} 237-238 (1963).

\bibitem{Kerr-Texas}
Roy Kerr, ``Gravitational collapse and rotation'', published in: 
{\sl Quasi-stellar sources and gravitational collapse:
Including the proceedings of the First Texas Symposium on Relativistic
Astrophysics}, edited by Ivor Robinson, Alfred Schild, and E.L. Sch\"ucking
(University of Chicago Press, Chicago, 1965), pages 99--102.\\
The conference was held in Austin, Texas, on 16--18 December 1963.



\bibitem{Einstein} 
Albert Einstein,
``Zur allgemeinen Relativitatstheorie'',
Sitzungsberichte der K\"oniglich Preussischen Akademie der Wissenschaften (1915)  778, 
Addendum-ibid.  (1915) 799. 


\bibitem{Hilbert} 
David Hilbert, ``Die Grundlagen der Physik (Erste Mitteilung)'', 
Nachrichten von der Gesellschaft der Wissenschaften zu G\"ottingen. 
Mathematisch-physikalische Klasse (1915), 395--407.


\bibitem{Schwarzschild1} 
Karl Schwarzschild, 
``\"Uber das Gravitationsfeld eines Massenpunktes nach der Einsteinschen Theorie'',
Sitzungsberichte der K\"oniglich Preussischen Akademie der Wissenschaften, 1916 vol. I, 189--196. 

\bibitem{Birkhoff}
Garret Birkhoff,  
 \emph{Relativity and Modern Physics}, 
(Harvard University Press, Cambridge, 1923).

\bibitem{Jebsen}
J\o{}rg Tofte Jebsen, ``\"Uber die allgemeinen kugelsymmetrischen 
L\"osungen der Einsteinschen Gravitationsgleichungen im Vakuum'', 
Ark. Mat. Ast. Fys. (Stockholm) {\bf 15} (1921) nr.18. 

\bibitem{Deser}
  Stanley Deser and Joel Franklin,
  ``Schwarzschild and Birkhoff \emph{a la} Weyl'',
  Am.\ J.\ Phys.\  {\bf 73} (2005) 261
  [arXiv:gr-qc/0408067].
  
 \bibitem{Ravndal} 
   Nils Voje Johansen, Finn Ravndal, 
   ``On the discovery of Birkhoff's theorem'', 
   Gen.Rel.Grav. 38 (2006) 537-540  [arXiv: physics/0508163].
   



\bibitem{Schwarzschild2} 
Karl Schwarzschild, 
``\"Uber das Gravitationsfeld einer Kugel aus inkompressibler Flussigkeit nach der Einsteinschen Theorie'',
Sitzungsberichte der K\"oniglich Preussischen Akademie der Wissenschaften, 1916 vol. I, 424--434. 



\bibitem{Oppenheimer}
J. Robert Oppenheimer and Hartland  Snyder,
``On Continued Gravitational Contraction'',
Phys. Rev. {\bf 56} (1939) 455.

\bibitem{Lense-Thirring}
Hans Thirring and Josef Lense,  ``\"Uber den Einfluss der Eigenrotation der Zentralk\"orperauf die Bewegung 
der Planeten und Monde nach der Einsteinschen Gravitationstheorie'', Physikalische Zeitschrift, Leipzig Jg. {\bf 19}  (1918), No. 8, p. 156--163.\\ 
English translation by Bahram Mashoon, Friedrich W. Hehl, and Dietmar S. Theiss, ``On the influence of the proper rotations of central bodies on the motions of planets and moons in Einstein's theory of gravity'', General Relativity and Gravitation  {\bf 16} (1984) 727--741.

\bibitem{Pfister}
Herbert Pfister, ``On the history of the so-called Lense--Thirring effect'',
{\sf http://philsci-archive.pitt.edu/archive/00002681/01/lense.pdf}

\bibitem{Adler-Bazin-Schiffer}
Ronald J. Adler, Maurice Bazin, and Menahem Schiffer, 
 {\sl Introduction to General Relativity}, Second edition, (McGraw--Hill, New York, 1975).\\
 {}[It is important to acquire the 1975 second edition, the 1965 first edition does not contain any discussion of the Kerr spacetime.]

\bibitem{MTW}
Charles Misner, Kip Thorne, and John Archibald Wheeler,  {\sl Gravitation}, 
(Freeman, San Francisco, 1973).

\bibitem{D'Inverno}
Ray D'Inverno, {\sl Introducing Einstein's Relativity}, (Oxford University Press, 1992).

\bibitem{Hartle}
James Hartle, {\sl Gravity: An introduction to Einstein's general relativity},
(Addison Wesley, San Francisco, 2003).

\bibitem{Carroll}
Sean Carroll, {\sl  An introduction to general relativity: Spacetime and Geometry},
(Addison Wesley, San Francisco, 2004).


 \bibitem{Chandrasekhar}
Subrahmanyan Chandrasekhar,  {\sl The Mathematical Theory of Black Holes}, 
(Oxford University Press, 1998).


\bibitem{Kerr1}
Roy Kerr, ``The Lorentz-covariant approximation method in general relativity''
 Il Nuovo Cimento. {\bf 13} (1959) 469.


\bibitem{Kerr2}
Roy Kerr and Joshua Goldberg,  
``Einstein Spaces With Four-Parameter Holonomy Groups'',
Journal of Mathematical Physics  {\bf 2} (1961) 332--336.


\bibitem{Kerr3}
Joshua Goldberg and Roy Kerr,
``Asymptotic Properties of the Electromagnetic Field'',
Journal of Mathematical Physics {\bf 5}  (1964) 172--176.



\bibitem{Kerr4}
Roy  Kerr, Alfred Schild,   
``Some algebraically degenerate solutions of Einstein's gravitational field equations'',
Proc. Symp. Appl. Math, 1965.
 
\bibitem{Kerr5} 
George Debney Jr, Roy Kerr, and Alfred Schild, 
``Solutions of the Einstein and Einstein-Maxwell Equations'', 
Journal of Mathematical
Physics {\bf 10}, 1842 (1969). 
 
\bibitem{Kerr6} 
Roy Kerr, George Debney Jr,  ``Einstein Spaces with Symmetry Groups'',
Journal of Mathematical Physics, {\bf 11} (1970) 2807--2817.


\bibitem{Kerr7}
Graham Weir and Roy Kerr, ``Diverging type-D metrics'',
Proceedings of the Royal Society of London, {\bf 355} (1977) 31--52.
 

\bibitem{Kerr8}
Roy Kerr and Wilson,~W. B., ``Singularities in the Kerr-Schild metrics'',  
General Relativity and Gravitation -- GR8 1977, proceedings of the 8th International Conference on General Relativity and Gravitation, held August 7-12, 1977, in Waterloo, Ontario, Canada. 1977, p. 378

 
\bibitem{Kerr9}
 Edward Fackerell and Roy Kerr, 
 ``Einstein vacuum field equations with a single non-null Killing vector'', 
 General Relativity and Gravitation,  {\bf 23} (1991) 861--878. 


\bibitem{Kerr10}
  Alexander Burinskii and Roy Kerr,
  ``Nonstationary Kerr congruences'',
  arXiv:gr-qc/9501012.
  





\bibitem{O'Neill} Barrett O'Niell, {\sl Geometry of Kerr Black Holes}, (A K Peters, 1995).

\bibitem{Hawking-Ellis}
Stephen Hawking and George Ellis, {\sl The large scale structure of space-time},
(Cambridge University Press, 1975). 



 \bibitem{Plebanski}
Jerzy Pleba\'nski and Andrzej Krasi\'nski, {\sl An introduction to general relativity and cosmology}, (Cambridge University Press, 2006). 


\bibitem{exact} Hans Stephani, Dietrich Kramer, Malcolm MacCallum, 
 Cornelius Hoenselaers, and Eduard Herlt,
  {\sl Exact Solutions of Einstein's Field Equations}, 
 (Cambridge University Press, 2002). 
 
 
\bibitem{vortex}
Matt~Visser and Silke~Weinfurtner,
  ``Vortex geometry for the equatorial slice of the Kerr black hole'',
  Class.\ Quant.\ Grav.\  {\bf 22} (2005) 2493
  [arXiv:gr-qc/0409014].

 \bibitem{Lake-invariants}
 K.~Lake,
  ``Differential Invariants of the Kerr Vacuum,''
  Gen.\ Rel.\ Grav.\  {\bf 36} (2004) 1159
  [arXiv:gr-qc/0308038].
  \\
 K.~Lake,
  ``Comment on negative squares of the Weyl tensor,''
  Gen.\ Rel.\ Grav.\  {\bf 35} (2003) 2271
  [arXiv:gr-qc/0302087].
 

 
 \bibitem{Doran}
Chris Doran,
  ``A new form of the Kerr solution'',
  Phys.\ Rev.\  D {\bf 61} (2000) 067503
  [arXiv:gr-qc/9910099].

  
\bibitem{Painleve}
Paul Painlev\'e,  ``La m\'ecanique classique et la theorie de la relativit\'e'', 
C. R. Acad. Sci., {\bf 173}, 677-680, (1921).
  
\bibitem{Gullstrand}
Allvar Gullstrand, ``Allgemeine L\"osung des statischen Eink\"orper\-problems in der Einsteinschen Gravitationstheorie'', 
Ark. Mat. Astron. Fys., {\bf 16}(8), 1-15, (1922).
  
\bibitem{analogue0}
Matt~Visser,
``Acoustic black holes: Horizons, ergospheres, and Hawking radiation'',
Class.\ Quant.\ Grav.\  {\bf 15}, 1767 (1998)
 [arXiv:gr-qc/9712010].


\bibitem{Smarr}
Lary Smarr, ``Surface geometry of charged rotating black holes'',
Physical Review D {\bf 7} (1973) 269--295.


\bibitem{Lake}
  Nikos Pelavas, Nicholas Neary and Kayll Lake,
  ``Properties of the instantaneous ergo surface of a Kerr black hole'',
  Class.\ Quant.\ Grav.\  {\bf 18} (2001) 1319
  [arXiv:gr-qc/0012052].
  
 
 \bibitem{mass-inflation}
 Eric Poisson and Werner Israel,
 ``Inner-horizon instability and mass inflation in black holes'',
  Phys. Rev. Lett. {\bf 63}, 1663--1666 (1989).
  \\
Roberto Balbinot and Eric Poisson,
``Mass inflation: The semiclassical regime'',
Phys. Rev. Lett. {\bf 70}, 13--16 (1993).

 
%
\end{thebibliography}
\end{document}